\documentclass[%
 reprint,
 superscriptaddress,
 footinbib,
 amsmath,amssymb,
 aps,
 prl,
 letterpaper,
]{revtex4-1}
\usepackage{graphicx}
\usepackage{dcolumn}
\usepackage{bm}
\usepackage[colorlinks = true]{hyperref}
\usepackage[dvipsnames]{xcolor}

\usepackage{mathtools}
\usepackage{siunitx}

\begin{document}

\title{Competing Time Scales Lead to Oscillations in Shear-Thickening Suspensions}

\author{J. A. Richards}
\email{jamesrichards92@gmail.com}
\author{J. R. Royer}
\affiliation{SUPA, School of Physics and Astronomy, The University of Edinburgh, King's Buildings, Peter Guthrie Tait Road, Edinburgh EH9 3FD, United Kingdom}
\author{B. Liebchen}
\affiliation{SUPA, School of Physics and Astronomy, The University of Edinburgh, King's Buildings, Peter Guthrie Tait Road, Edinburgh EH9 3FD, United Kingdom}

\affiliation{Institut f{\"u}r Theoretische Physik II: Weiche Materie,
Heinrich-Heine-Universit{\"a}t D{\"u}sseldorf, D-40225 D{\"u}sseldorf, Germany}
\author{B. M. Guy}
\author{W. C. K. Poon}
\affiliation{SUPA, School of Physics and Astronomy, The University of Edinburgh, King's Buildings, Peter Guthrie Tait Road, Edinburgh EH9 3FD, United Kingdom}

\date{\today}

\begin{abstract}
Competing time scales generate novelty. Here, we show that a coupling between the time scales imposed by instrument inertia and the formation of inter-particle frictional contacts in shear-thickening suspensions leads to highly asymmetric shear-rate oscillations. Experiments tuning the presence of oscillations by varying the two time-scales support our model. The observed oscillations give access to a shear-jamming portion of the flow curve that is forbidden in conventional rheometry. Moreover, the oscillation frequency allows us to quantify an intrinsic relaxation time for particle contacts. The coupling of fast contact network dynamics to a slower system variable should be generic to many other areas of dense suspension flow, with instrument inertia providing a paradigmatic example. 
\end{abstract}

\maketitle

Concentrated suspensions of non-Brownian (or granular) particles in a Newtonian solvent occur widely in industry, \textit{e.g.}, concrete \cite{van2018concrete}  mine tailings \cite{boger2009rheology}, and chocolate~\cite{afoakwa2007factors}. Their viscosity, $\eta$, often increases with either shear rate, $\dot\gamma$, or stress, $\sigma$ \cite{denn2014rheology}. Such shear thickening is now understood as a transition from a low-viscosity state, with lubricated particle contacts, to a high-viscosity state, with frictional contacts, as the repulsive force between particles is overcome at a critical onset stress, $\sigma^*$ \cite{mari2015nonmonotonic, lin2015hydrodynamic, comtet2017pairwise, clavaud2017revealing}. 

A phenomenological model of this process by Wyart and Cates (WC) \cite{wyart2014discontinuous} predicts three types of flow curve, $\sigma(\dot{\gamma})$. At low volume fraction, $\phi$, a smooth increase connects two constant-slope (= viscosity) branches in $\sigma(\dot\gamma)$, giving rise to continuous shear thickening (CST). Above a critical $\phi_{\rm DST}$, $\sigma(\dot\gamma)$ becomes \textsf{S}-shaped, with a backwards-bending (d$\sigma$/d$\dot{\gamma}<0$) region connecting the two branches, giving discontinuous shear thickening (DST). Finally, above some $\phi_{m}$, $\sigma(\dot\gamma)$ has no flowing upper branch and it bends back to $\dot{\gamma}=0$: the system shear jams at high stresses.

In the CST regime, suspensions flow steadily and homogeneously, and the WC model fits data from nearly-monodisperse hard-sphere systems  \cite{guy2015towards, royer2016rheological}. In the DST regime, there is a jump in $\sigma$ as the imposed $\dot{\gamma}$ is increased \cite{bender1996reversible}, while under imposed $\sigma$, neither homogeneous nor shear-banded steady flow is possible \cite{hermes2016unsteady}. There is no general model for the system-specific flow in this regime. Recent experiments \cite{saint2018uncovering,rathee2017localized} and simulations \cite{chacko2018dynamic} focus on banding: spatial variation with high-$\sigma$ and low-$\sigma$ regions. Many systems also show large temporal fluctuations \cite{lootens2003giant}, which sometimes begin as `relaxation oscillations': $\dot\gamma$ periodically drops precipitously to a nearly-jammed state \cite{larsen2014fluctuations, bossis2017discontinuous, bossis2018tunable}, with a frequency that increases with applied stress \cite{nagahiro2013experimental, hermes2016unsteady}. We extend the WC model to account quantitatively for such oscillations.

The key physics is the competition between the  dynamics of frictional contact formation and a `system variable', here the acceleration of the rheometer geometry \cite{larsen2014fluctuations, bossis2017discontinuous}. When the ratio of the time scale of the former to that of the latter is small, we predict homogeneous flow with relaxation-type $\dot{\gamma}$ oscillations. Fitting the observed $\sigma$-dependence of the oscillation frequency reveals and quantifies an additional time scale, that intrinsic to the relaxation of frictional contacts after formation. Thus, rheometer geometry inertia, often considered an artefact, can be used to probe suspensions near jamming.

\begin{figure}
    \centering
    \includegraphics[trim=0 0 0 0, clip, width=\columnwidth]{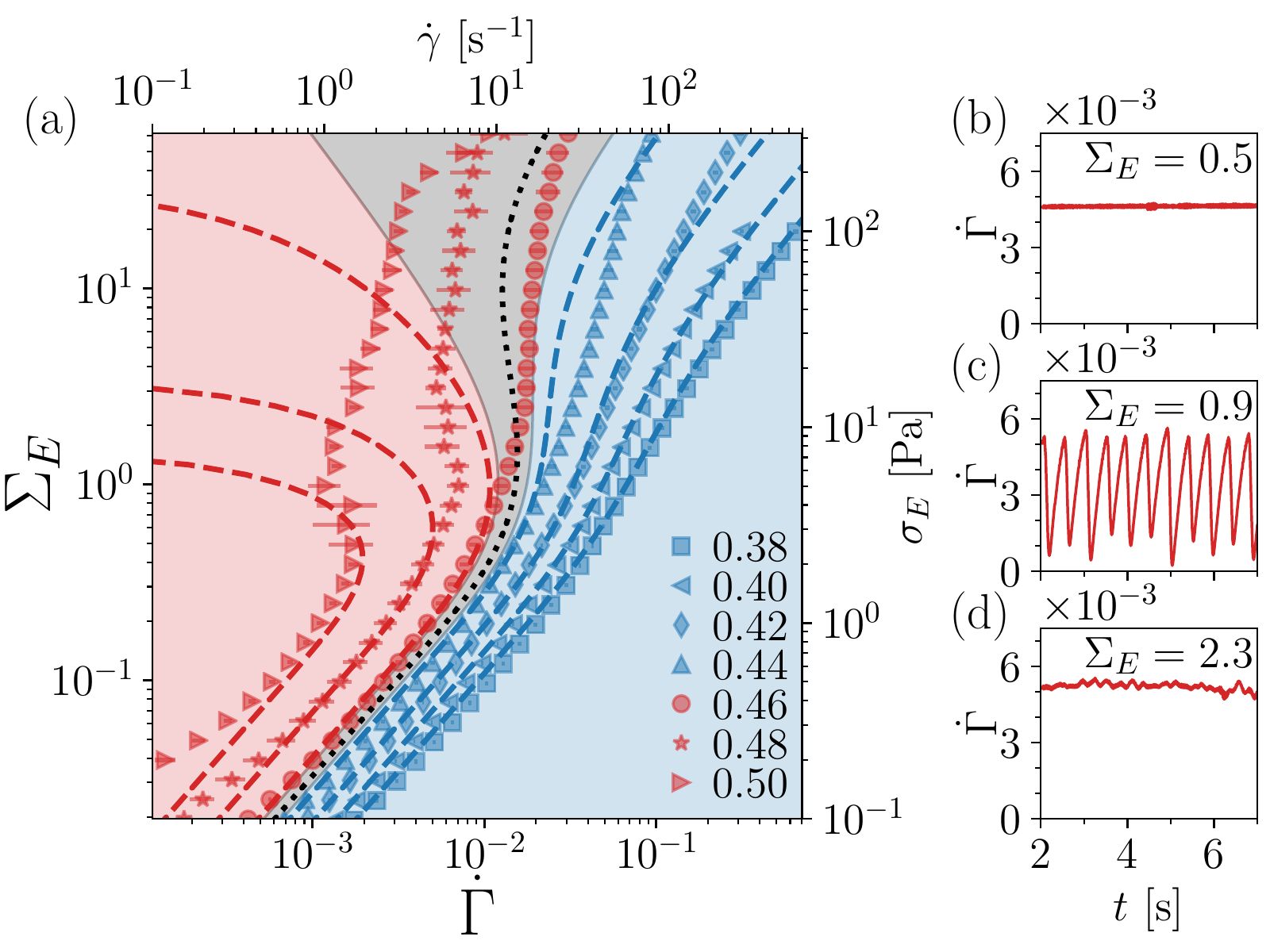}\\
    \includegraphics[trim=0 0 0 0, clip, width=\columnwidth]{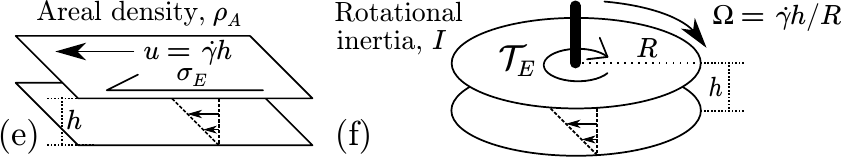}
    \caption{Imposed-stress rheology. (a) Flow curves: dimensionless stress {\it vs.}~dimensionless average shear rate, $\Sigma_{E}(\dot\Gamma)$, and absolute, $\sigma_{E}(\dot{\gamma})$, at given weight fractions, $\varphi$.
    Dashed lines: WC model for $\varphi_{m}=0.457$, $\varphi_{\rm rcp}=0.546,~\beta=0.94$ ($\Rightarrow \varphi_{\rm DST}=0.445$) and $\sigma^*=\SI{5.1}{\pascal}$ at given $\varphi$ (- -) and $\varphi=0.45$ ($\cdot \cdot$). Blue, $\varphi < \varphi_{\rm DST}$; grey, $\varphi_{\rm DST} < \varphi < \varphi_{m}$; red, $\varphi>\varphi_{m}$. Symbols: $\Sigma_{E}$ \textit{vs.} time-averaged $\dot{\Gamma}$ for cornstarch suspensions in 50~wt.\% glycerol-water. Error bars denote standard deviation from three up-sweeps. (b)-(d) Time-dependent experimental shear rate, $\dot{\Gamma}(t)$, for $\varphi=0.48$, showing respectively: (b) steady flow below the onset of shear thickening, (c) periodic shear-rate oscillations and (d) aperiodic flow at high stress. (e) Rheometric geometry: infinite plates, separation $h$, velocity $u$ ($\dot{\gamma}=u/h$), areal density $\rho_{A}$ and applied stress $\sigma_E$. (f) Experimental geometry: rotating plates, radius $R$, gap height $h$, relative angular velocity $\Omega$ ($\dot{\gamma}=\Omega R / h$), rotational inertia $I$ and applied torque $\mathcal{T}_E = \sigma_E \pi R^3 / 2$. Equivalent $\rho_A = 2I / \pi R^4 $, using Eq.~\ref{eq:dgdt}.}
    \label{fig:FC}
\end{figure}

WC introduced a stress-dependent steady-state fraction of frictional contacts, $\hat{f}$; simulations \cite{mari2014shear,chacko2018dynamic} find 
\begin{equation}
     \hat{f} ( \sigma ) = \exp \left ( - ( \sigma ^ * / \sigma ) ^ {\beta } \right ),
     \label{eq:fSS}
\end{equation} 
with $\beta \lesssim 1$. $\hat{f}$ controls the jamming point, $\varphi_{J}$, at which $\eta \to \infty$. Here we use weight fractions, $\varphi$, due to the porosity of our main model system (cornstarch) \cite{han2017measuring}. At $\sigma \ll \sigma^*$, $\hat{f} = 0$, and the system jams at random close packing, $\varphi_{J}=\varphi_{\rm rcp}$.  When $\sigma \gg \sigma^*$, $\hat{f} \to 1$ and the system jams at some $\varphi_{m}<\varphi_{\rm rcp}$. The WC model linearly interpolates between these two limits:
\begin{equation}
    \varphi _ { J } ( \hat{f} ) = \varphi _ { m } \hat{f} + \varphi _ { \rm rcp } ( 1 - \hat{f} ) .
    \label{eq:phiJ}
\end{equation}
The distance to jamming then determines $\eta$ via
\begin{equation}
    \eta(\varphi,\varphi_{J}) = \eta _ {s } \left [ 1 - \varphi /\varphi_{J }( \hat{f} )  \right ]^{-2},
    \label{eq:viscosity}
\end{equation}
with $\eta_{s}$ the solvent viscosity. At a given weight fraction, $\hat{f}$ increases with stress, lowering the jamming point, which in turn increases the viscosity, $\eta(\sigma) \equiv \eta \{ \varphi,\varphi_{J}[\hat{f}(\sigma)]\}$. At $\varphi_{\rm DST} < \varphi < \varphi_{m}$, the flow curve, $\sigma(\dot{\gamma})$, becomes $\mathsf{S}$-shaped, Fig.~\ref{fig:FC}(a) ($\cdot \cdot$), with a region where d$\sigma$/d$\dot{\gamma}<0$. At $\varphi>\varphi_{m}$, shear jamming (SJ) is predicted, with the flow curve doubling back to $\dot{\gamma}=0$ when $\varphi_{J}(\sigma)=\varphi$, Fig.~1(a) ({\color{red} - -}). 

In non-steady flow, {\it e.g.}~on reversal \cite{lin2015hydrodynamic,peters2016rheology}, the contact network of the suspension takes finite time to adapt. Thus, the fraction of frictional contacts at any one instant, $f(t)$, may differ from its steady-state value, $\hat f$, given by Eq.~\ref{eq:fSS}, towards which $f(t)$ relaxes. Simulations show that $f$ evolves with the accumulated strain \cite{mari2015nonmonotonic}; so following previous work \cite{mari2015nonmonotonic,nakanishi2012fluid}, we write \begin{equation}
    \frac { \mathrm{ d } f } { \mathrm { d } t } = - \frac{\dot { \gamma } } { \gamma _ { 0 } } \left [ f - \hat { f } ( \sigma ) \right ],
    \label{eq:dfdt}
\end{equation}
with a characteristic strain, $\gamma_0$ (and $\dot{\gamma} \geq 0$). We now use Eqs.~\ref{eq:phiJ} and \ref{eq:viscosity} to relate $\eta$ to $f$, rather than just $\hat{f}$.

External stress, $\sigma_{E}$, is applied through the system boundaries. In a rheometer, this is the `geometry', which has far higher mass than the suspension for a typical gap height, $h$, between the boundaries \cite{lauger2016effects}, Fig.~\ref{fig:FC}(e). In the steady state, $\sigma_{E}=\eta(\hat f)\dot{\gamma}$, the sample stress, $\sigma$. When $\mathrm{d} \dot{ \gamma } /  \mathrm{ d } t\neq 0$, force balance between the geometry and the sample gives
\begin{equation}
    \rho_{A} h \frac {\mathrm{d} \dot{ \gamma } } { \mathrm{ d } t } = \sigma_{E} - \eta(f)\dot{\gamma},
    \label{eq:dgdt}
\end{equation}
with $\rho_{A}$ the geometry's areal density. Equations~\ref{eq:dfdt} and \ref{eq:dgdt}, being two-dimensional, cannot capture aperiodic flow, but can account for $\dot\gamma$-oscillations and elucidate the physics of unsteady flow in shear-thickening suspensions.

Measuring time in units of the geometry inertial time scale, $t_{i} = \rho_{A}h/\eta_{s}$, we rewrite Eqs.~\ref{eq:dgdt} \& \ref{eq:dfdt} as:
\begin{align}
    \frac{\mathrm{d} \dot{\Gamma}}{\mathrm{d}\tau} = \Sigma_{E} - \eta_{r}(f)\dot{\Gamma}  \equiv g_1(\dot\Gamma,f) \label{eq:DG},\\
    \frac{\mathrm{d}f}{\mathrm{d}\tau} = - \frac{\dot{\Gamma}}{\epsilon} \left[ f - \hat{f}(\eta_{r}(f)\dot{\Gamma}) \right] \equiv g_2(\dot\Gamma,f), \label{eq:DF}
\end{align}
where $\tau = t/t_{i}$. Other dimensionless variables are shear rate, $\dot{\Gamma} \equiv \mathrm{d}\Gamma / \mathrm{d} \tau = \dot{\gamma}\eta_{s} / \sigma^*$; applied stress, $\Sigma_{E}=\sigma_E/\sigma^*$; viscosity, $\eta_{r}=\eta(f)/\eta_{s}$; sample stress, $\eta_{r}(f) \dot{\Gamma} = \eta(f)\dot{\gamma}/\sigma^*$; and, strain, $\Gamma = \eta_{s}^2/(\rho_{A}h\sigma^*)$.

The time scale for contact network formation, $t_{c}= \gamma_0 \eta_{s} / \sigma^*$, competes with the inertial time, yielding our key dimensionless parameter,
\begin{equation}
\epsilon = \frac{t_{c}}{t_{i}} \equiv \frac{\gamma_0 \eta_{s}^2}{ \rho_{A} h \sigma^*}. \label{eq:epsilon}
\end{equation}
When $t_{i} \gg t_{c}$, {\it i.e.} $\epsilon \ll 1$, Eqs.~\ref{eq:DG} and \ref{eq:DF} form a singular autonomous system \cite{hinch1991}, which may undergo a Hopf bifurcation to show relaxation oscillations as the control parameter $\Sigma_{E}$ is varied \cite{baer1986}.

\begin{figure}
    \centering
    \begin{minipage}{\columnwidth}
    \includegraphics[trim=0 0 0 0, clip, width=.49\textwidth]{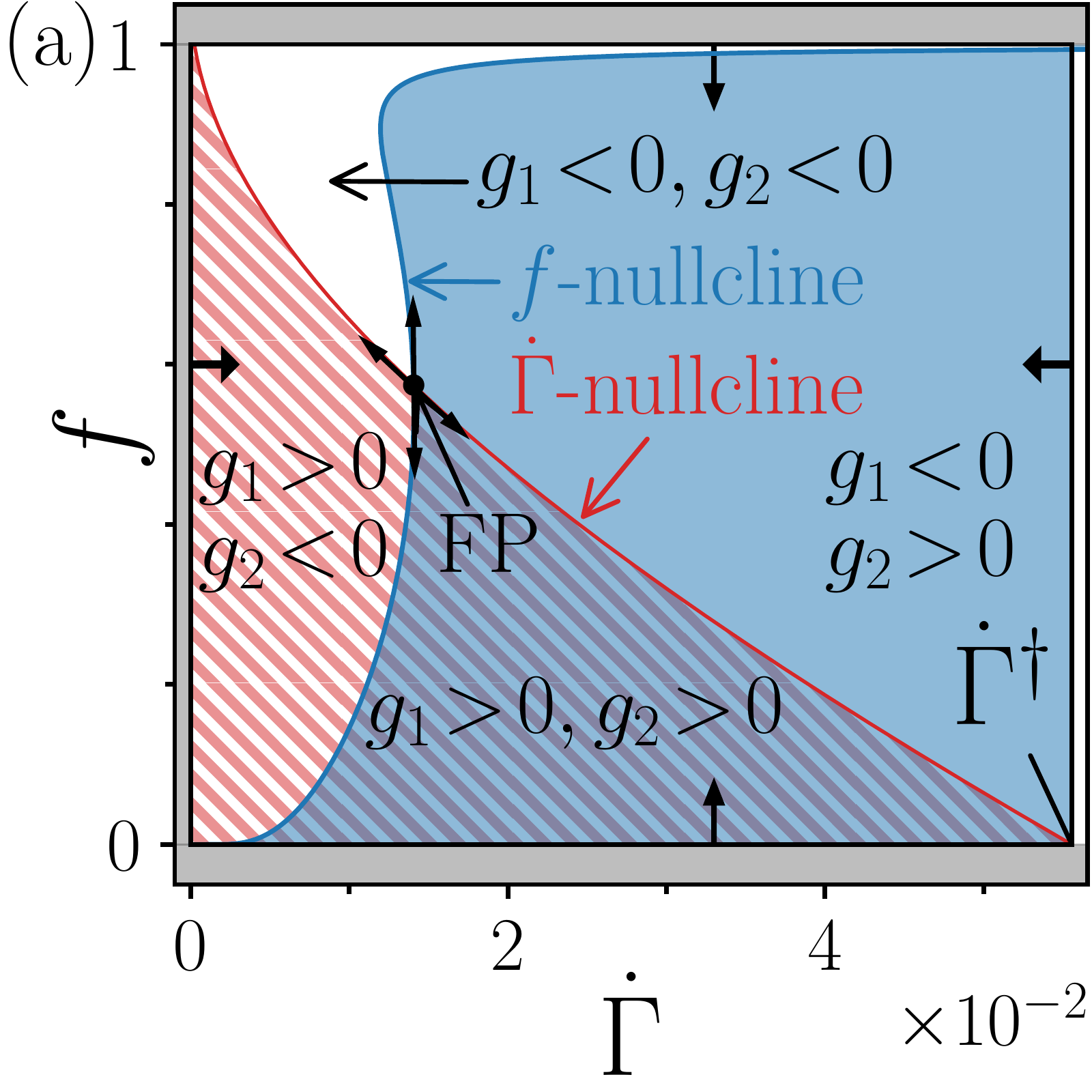}
    \includegraphics[trim=0 0 0 0, clip, width=.49\textwidth]{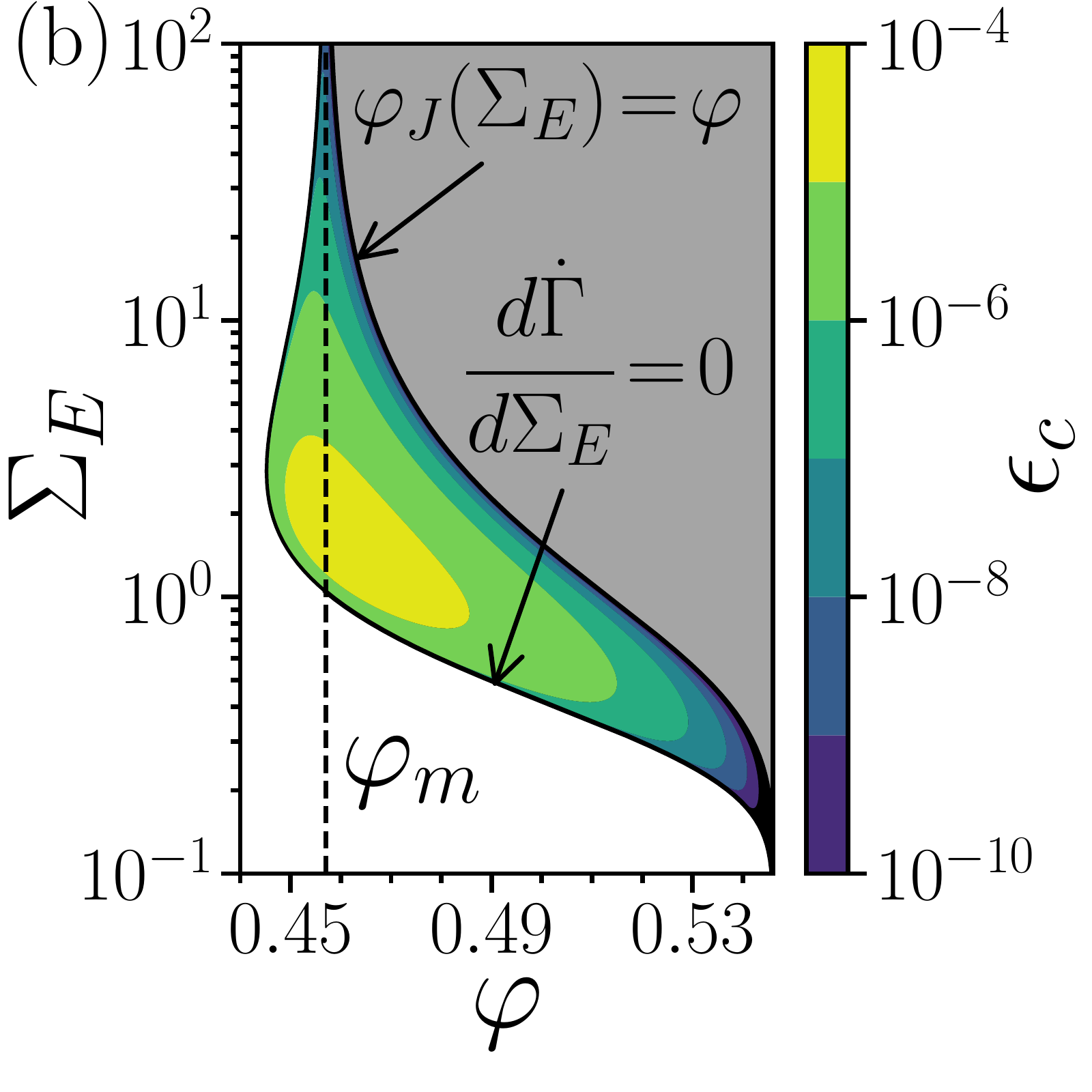}
    \end{minipage}
    \\
    \begin{minipage}{\columnwidth}
     \includegraphics[trim=0 0 0 0, clip, width=.49\textwidth]{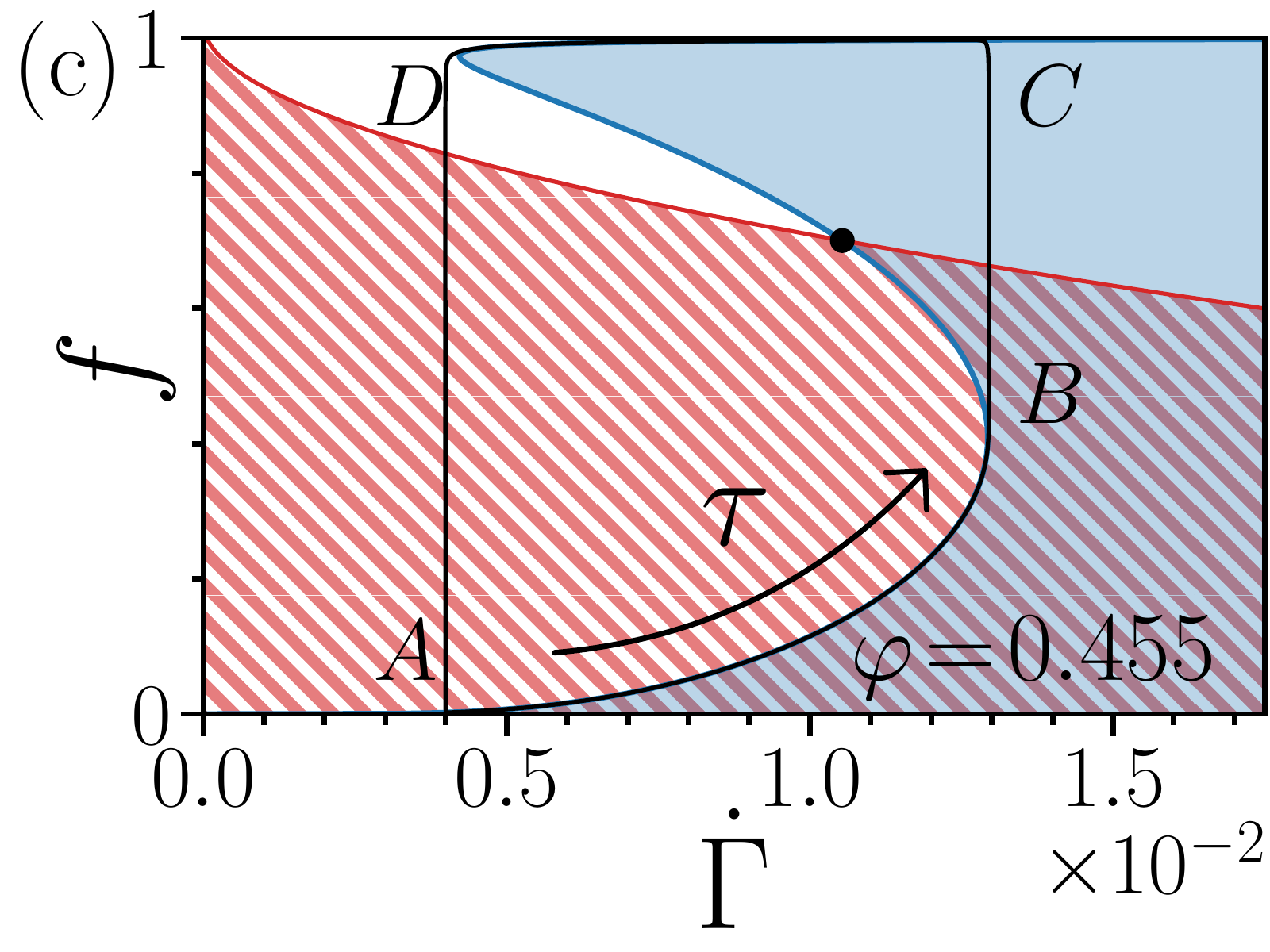}
     \includegraphics[trim=0 0 0 0, clip, width=.49\textwidth]{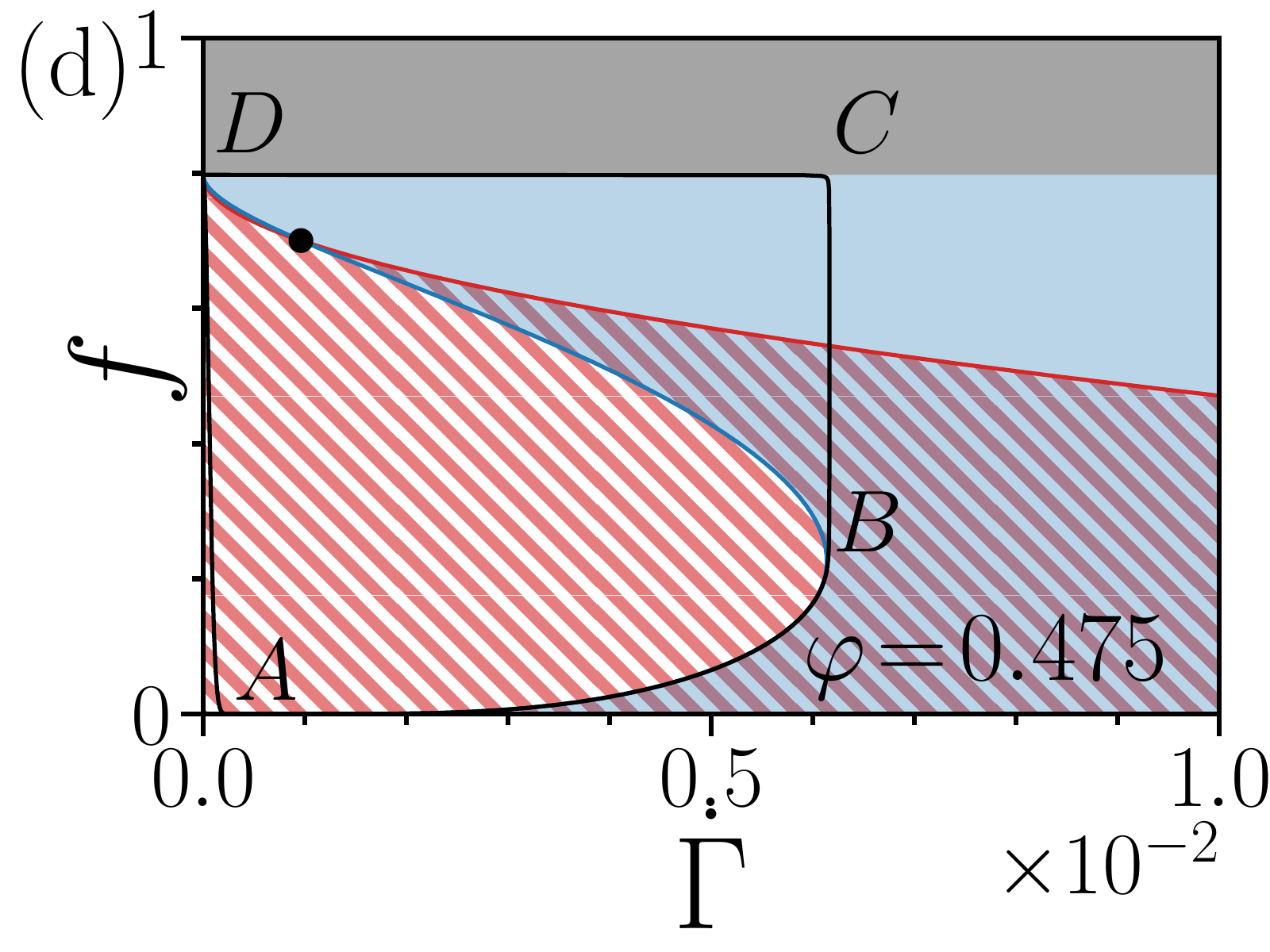}
    \end{minipage}
    \caption{Limit-cycle behavior for Eqs.~\ref{eq:DG} and \ref{eq:DF}. (a) Phase-plane schematic for $\varphi_{\rm DST} < \varphi < \varphi_{m}$. Red line, $\dot{\Gamma}$-nullcline; red hatched shading, $g_1>0$; blue line, $f$-nullcline; blue shading, $g_2>0$. Fixed point, FP. On the black rectangle, trajectories point inwards, indicating the existence of a limit cycle if FP is unstable. (b) Critical stability criterion value, $\epsilon_{c}$, from Eq.~\ref{eq:criterion}. Solid black lines $\epsilon_{c}=0$; grey shading, shear jammed. (c) Limit cycle for \textsf{S}-shaped flow curve; $\varphi = 0.455$ and $\Sigma_{E}=3.0$; WC model parameters from Fig.~\ref{fig:FC}. Black line, numerical solution for $\epsilon=10^{-9}$; shading as in (a). (d) Limit cycle for SJ flow curve; $\varphi = 0.475$ and $\Sigma_{E}=3.0$; grey shading, shear jammed; other parameters and shading as in (c).
    }
    \label{fig:analysis}
\end{figure}

For a given $\varphi$ and $\Sigma_{E}$, a fixed point occurs where the nullclines $g_1 = 0$ and $g_2 = 0$ intersect, Fig.~\ref{fig:analysis}(a). Analysing the Jacobian \cite{scheinerman1996}, $ \begin{psmallmatrix}\partial g_1/\partial\dot\Gamma & \partial g_1/\partial f\\ \partial g_2/\partial \dot\Gamma & \partial g_2/\partial f \end{psmallmatrix}$, shows that this fixed point is unstable if 
\begin{equation}
    \epsilon < \epsilon_{c} = - \dot{\Gamma}(\mathrm{d}\dot{\Gamma}/\mathrm{d}\Sigma_\mathrm{E}),
    \label{eq:criterion}
\end{equation}
which, since $\epsilon > 0$, requires d$ \dot{ \Gamma } $/d$ \Sigma_{E} < 0 $, {\it i.e.}~a backwards-bending flow curve, see Supplemental Material for derivation~\cite{SM}. Thus, the DST-boundary (d$\dot{\Gamma}$/d$\Sigma_{E}$=0) forms the lower boundary of our region of potential instability, Fig.~\ref{fig:analysis}(b). The upper boundary of this occurs at shear jamming, $\varphi_{J}(\Sigma_{E}) = \varphi$, where the flow curve touches the vertical axis so that $\dot\Gamma=0$.  Above this boundary, no flow is possible. Between these two boundaries, $\epsilon_{c}(\varphi,\Sigma_{E})$ peaks at $\epsilon_{c}^{\max} = 2\times10^{-5}$: instability may occur between DST and shear jamming whenever $\epsilon < 2\times10^{-5}$~\footnote{$\epsilon_c^{\max}$ is found from the maximum numerically calculated $\epsilon_c$ in Fig.~\ref{fig:analysis}(b) with $\delta \varphi \approx 10^{-5}$ and $\delta \log \Sigma_E = 3 \times 10^{-4}$.}. Physically, at such small $\epsilon$  ({\it i.e.}~$t_{i} \gg t_{c}$), the suspension thickens before the geometry slows, so the sample stress rises, driving $\hat{f}$ higher and causing further thickening in a vicious cycle, pushing the system away from the steady state. 

We now describe our dynamical system by phase-plane trajectories that depend parametrically on $\tau$. Consider the regime $\varphi_{\rm DST} \leq \varphi < \varphi_{m}$ with \textsf{S}-shaped flow curves, Fig.~1(a) (- -). The $f$-nullcline, Fig.~\ref{fig:analysis}(a), reflects the shape of the steady-state flow curve \footnote{As there is a one-to-one dependence of $\Sigma_{E}$ to $f$ on the $f$-nullcline, $g_2 = 0$, see Eq.~\ref{eq:DF}.}. Equations \ref{eq:DG} and \ref{eq:DF} show that trajectories point {\it inwards} everywhere on the rectangle defined by $\dot\Gamma = 0$, $\dot\Gamma = \dot\Gamma^\dagger$ (where the $\dot\Gamma$-nullcline intersects the $\dot\Gamma$ axis), $f = 0$ and $f = 1$, Fig.~\ref{fig:analysis}(a). However, trajectories point {\it outwards} on any infinitesimally-small loop around the fixed point if it is unstable. The Poincar\'e-Bendixson Theorem \cite{scheinerman1996} then predicts a limit cycle in the region depicted in Fig.~\ref{fig:analysis}(a) if $\epsilon < \epsilon_{c}$. 

A numerically-calculated limit cycle after the onset of DST is shown in Fig.~\ref{fig:analysis}(c). To understand this cycle, divide Eq.~\ref{eq:DF} by Eq.~\ref{eq:DG} to obtain
\begin{equation}
    \epsilon \left[ \frac{\Sigma_{E}}{\dot{\Gamma} } - \eta_{r}(f)\right] = - \left[ f - \hat{f}(\eta_{r}(f)\dot{\Gamma}) \right] \frac{\mathrm{d}\dot{\Gamma}}{\mathrm{d}f}.
    \label{eq:phase}
\end{equation}
If $\epsilon \rightarrow 0$, Eq.~\ref{eq:phase} requires d$\dot\Gamma/$d$f \to 0$ (vertical lines) or $ f \rightarrow \hat{f}(\eta_r \dot{\Gamma})$ ($f$-nullcline). If $\epsilon \ll 1$, starting at $(0,0)$, the system follows the $f$-nullcline ($g_2=0,~g_1>0$), Fig.~\ref{fig:analysis}(c), at a rate controlled by $t_{i}$ (Eq.~\ref{eq:DG}). At $B$, the system jumps vertically to join the `upper branch' of the $f$-nullcline at $C$.  It now follows the `upper branch' of the $f$-nullcline ($g_2=0,~g_1<0$) until it reaches $D$, where it drops vertically to $A$, and the process repeats: we have a limit cycle. As a consistency check, the `jump' $BC$ and hence the limit cycle relies on $\dot{\gamma}$ not changing ($t \ll t_{i}$) as a large number of frictional contacts form and the suspension shear thickens ($t > t_{c}$), {\it i.e.} $\epsilon \ll 1$, as assumed. 

At $\varphi>\varphi_{m}$, Fig.~\ref{fig:analysis}(d), the `jump' from $B$ takes the suspension towards jamming, $\eta \to \infty$ at $C$, whereupon $\dot\gamma$ abruptly goes to zero, giving a horizontal `jump' to $D$, from where the system drops back to $A$ on the $f$-nullcline, again giving a limit cycle. Note that the $CD$ part of our limit cycle probes our system close to jamming. Unlike in conventional steady-state rheology \cite{fall2010shear}, our system should remain homogeneous: the time needed to traverse $BCD$ is simply too short to allow finite particle migration.

\begin{figure}[t]
    \centering
    \includegraphics[width=\columnwidth,trim=0 0 0 0, clip]{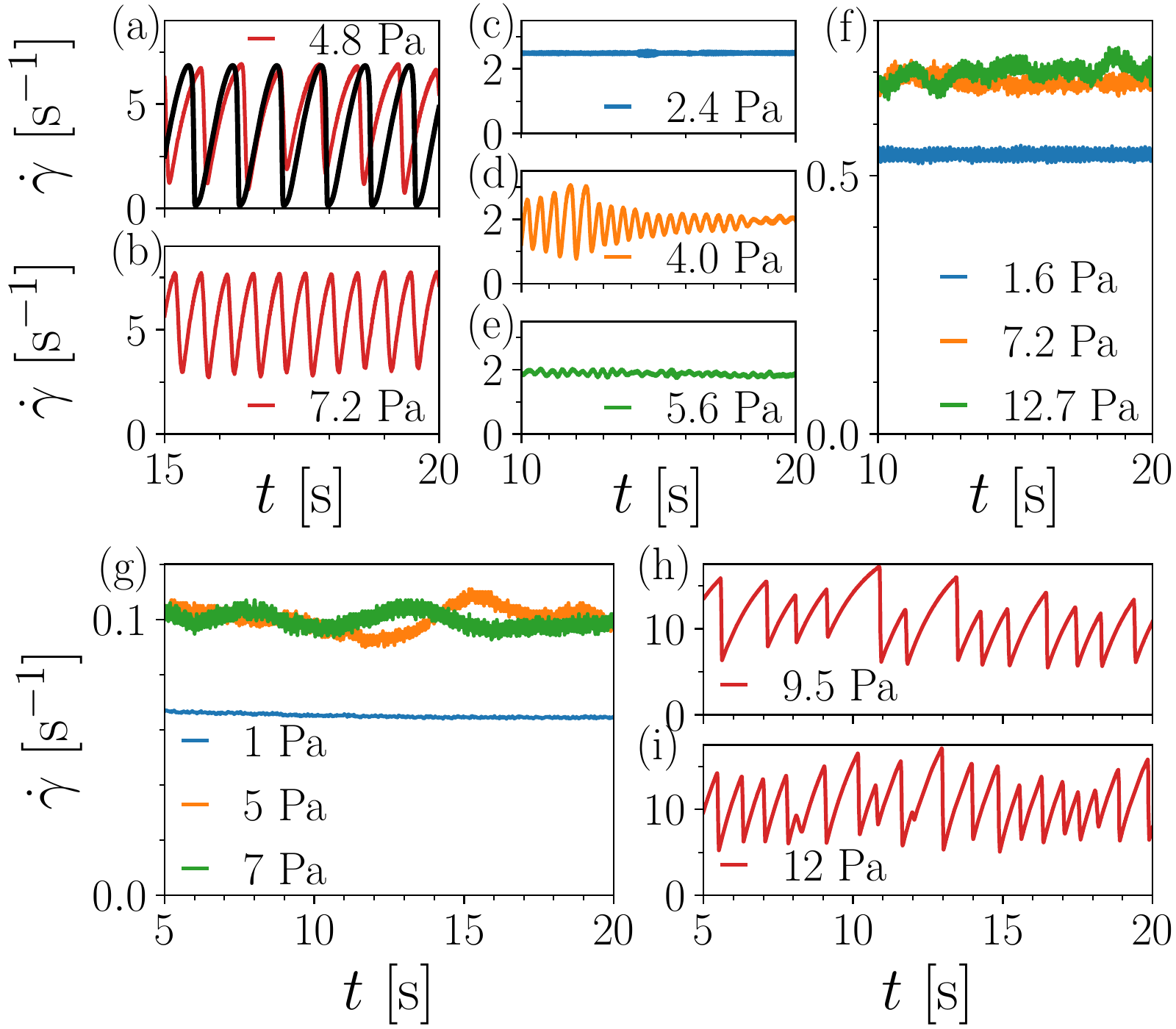}
    \caption{Tuning shear-rate oscillations for $\varphi \gtrsim \varphi_{\rm DST}$. (a)-(b) Low viscosity: cornstarch in 50~wt.\% glycerol-water, $\varphi=0.47$; relaxation oscillations. Red, experimental data; black, model: $\epsilon = 2.7 \times 10^{-6}~$, flow curve parameters from Fig.~\ref{fig:analysis}. Traces aligned by eye. (c)-(e) Medium viscosity: cornstarch in 67~wt.\% glycerol-water, $\eta_{s}=\SI{15}{\milli \pascal \second}$ at $\varphi=0.45 \gtrsim \varphi_{\rm DST} \approx 0.44$; damped oscillations in a narrow range of stress. (f) High viscosity: cornstarch in 85~wt.\% glycerol-water, $\varphi=0.44 \gtrsim \varphi_{\rm DST}\approx 0.44$; DST with no relaxation oscillations. (g) \SI{4}{\micro \meter} silica spheres in 87~wt.\% glycerol-water ($\eta_{s}=\SI{151}{\milli \pascal \second}$ and $\epsilon = 1 \times 10^{-4} $) at $\phi=0.574 \gtrsim \phi_{m}=0.57$; DST with no large shear-rate oscillations. (h)-(i) Silica in dimethyl sulfoxide-water mixture ($\eta_{s}=\SI{3.4}{\milli \pascal \second}$ and $\epsilon = 2 \times 10^{-7}$) at $\phi=0.58$; shear-rate oscillations.}
    \label{fig:trace}
\end{figure}

To validate our model, we first characterized a shear-thickening suspension known to show oscillations~\cite{hermes2016unsteady}. Cornstarch (Sigma Aldrich, particle diameter $\approx \SI{14}{\micro\meter}$ and polydispersity $\approx$ 40\% from static light scattering~\cite{hermes2016unsteady}) was dispersed into 50~wt.\% glycerol-water ($\eta_{s} = \SI{6}{mPa.s}$). We used a TA Instruments DHR-2 with roughened parallel plates (radius $R = \SI{20}{\milli m}$ and $h = \SI{1.0}{\milli m}$ for flow curves, \SI{1.5}{mm} for time dependence), Fig.~1(f). Rim shear rates, $\dot{\gamma}=\Omega R/h$, and apparent stresses, $\sigma_E = 2\mathcal{T}_E/\pi R^3$, come from the applied torque, $\mathcal{T}_E$, and measured angular velocity, $\Omega$. Cornstarch particles are porous~\cite{han2017measuring}; so we quote weight fractions, $\varphi$, using freshly-prepared samples and monitoring reproducibility. 

The WC model captures credibly the time-averaged flow curves of this system for $\varphi < \varphi_{\rm DST}$, Fig.~\ref{fig:FC}(a) \mbox{(\textcolor{blue}{-~-})}, with $\varphi_{m}$, $\varphi_{\rm rcp}$, $\sigma^*$ and $\beta$ determined from fitting the steady-state rheology (see SM~\cite{SM}), averaging 3 upsweeps at 10pts./decade from \SI{0.1}{\pascal} to fracture with \SI{10}{\second} average and \SI{5}{\second} delay, separately ensuring reversibility. At $\varphi > \varphi_{m}$, the WC model works until the predicted flow curve bends backwards, Fig.~1(a) ({\color{red} - -}). Up to this point, the flow is steady: $\dot\gamma$ is constant in time, Fig.~\ref{fig:FC}(b). At higher stress, the flow starts to oscillate, Fig.~\ref{fig:FC}(c), before becoming aperiodic, Fig.~\ref{fig:FC}(d)~\cite{hermes2016unsteady}.

The measured geometry moment of inertia, $I$, gave $\rho_{A}\equiv 2 I/\pi R^4 = \SI{175}{\kilo \gram \per \meter \squared}$~\cite{SM}. Imposed-rate experiments gave $\gamma_0 = \mathcal{O}(10^{-1})$~\cite{SM}. Thus, $t_{c} = \SI{1.1D-4}{\second}$, $t_{i} = \SI{44}{\second}$, and $\epsilon= 2.7\times10^{-6}$, far below the $\epsilon_{c}^{\max} = 2\times10^{-5}$ for observing instability when $\varphi_{\rm DST} < \varphi < \varphi_{\rm rcp}$. Solving Eqs.~\ref{eq:DG} and \ref{eq:DF} numerically at $\varphi = 0.47$ and $\Sigma_{E} = 0.93$, we find relaxation oscillations quantitatively matching experiments with no free parameters, Fig.~\ref{fig:trace}(a). 

Next, we varied $\epsilon \propto \eta_{s}^2$ by increasing the solvent glycerol proportion, see SM for time-averaged rheology \cite{SM}. For $\eta_{s} = \SI{15}{\milli \pascal \second}$, $ \epsilon \sim 2 \times 10^{-5} \gtrsim \epsilon_{c}^{\max}$, only damped oscillations in a narrow stress range were observed, Fig.~\ref{fig:trace}(c)-(e). For $\eta_{s} = \SI{75}{\milli\pascal\second}$, $\epsilon \sim 3\times 10^{-4} \gg \epsilon_{c}^{\max}$, no shear-rate oscillations are seen at stresses and weight fractions in the DST-regime, Fig.~\ref{fig:trace}(f). Oscillations could also be eliminated by only reducing $h$ (increasing $\epsilon \gtrsim \epsilon_{c}^{\max}$, see SM~\cite{SM}), however large variation of $\rho_A$ is restricted by rheometer design. We also studied shear-thickening silica suspensions (diameter \SI{4}{\micro\meter})~\cite{royer2016rheological}, in which oscillations have not been reported before. Experiments were performed using an Anton-Paar MCR302 in a parallel-plate geometry ($R=\SI{20}{\milli \metre},~h=\SI{1.5}{\milli \metre}$) with $\rho_{A}=\SI{400}{\kilo \gram \meter^{-2}}$, see SM for details \cite{SM}. In 87~wt.\% glycerol-water with $\eta_{s} = \SI{151}{\milli \pascal \second}$ and $\epsilon = 1 \times 10^{-4}$, no oscillations were seen, Fig.~\ref{fig:trace}(g). Reducing $\eta_{s}$ to \SI{3.4}{\milli \pascal \second} using a dimethyl sulfoxide-water mixture, giving $\epsilon = 2 \times 10^{-7} $, we found relaxation oscillations, Figs.~\ref{fig:trace}(h) and \ref{fig:trace}(i). All our available data are consistent with the predicted $\epsilon_{c}^{\max} = 2 \times 10^{-5}$ for instability.

\begin{figure}[t]
    \centering
    \includegraphics[width=\columnwidth,trim=0 0 0 0, clip]{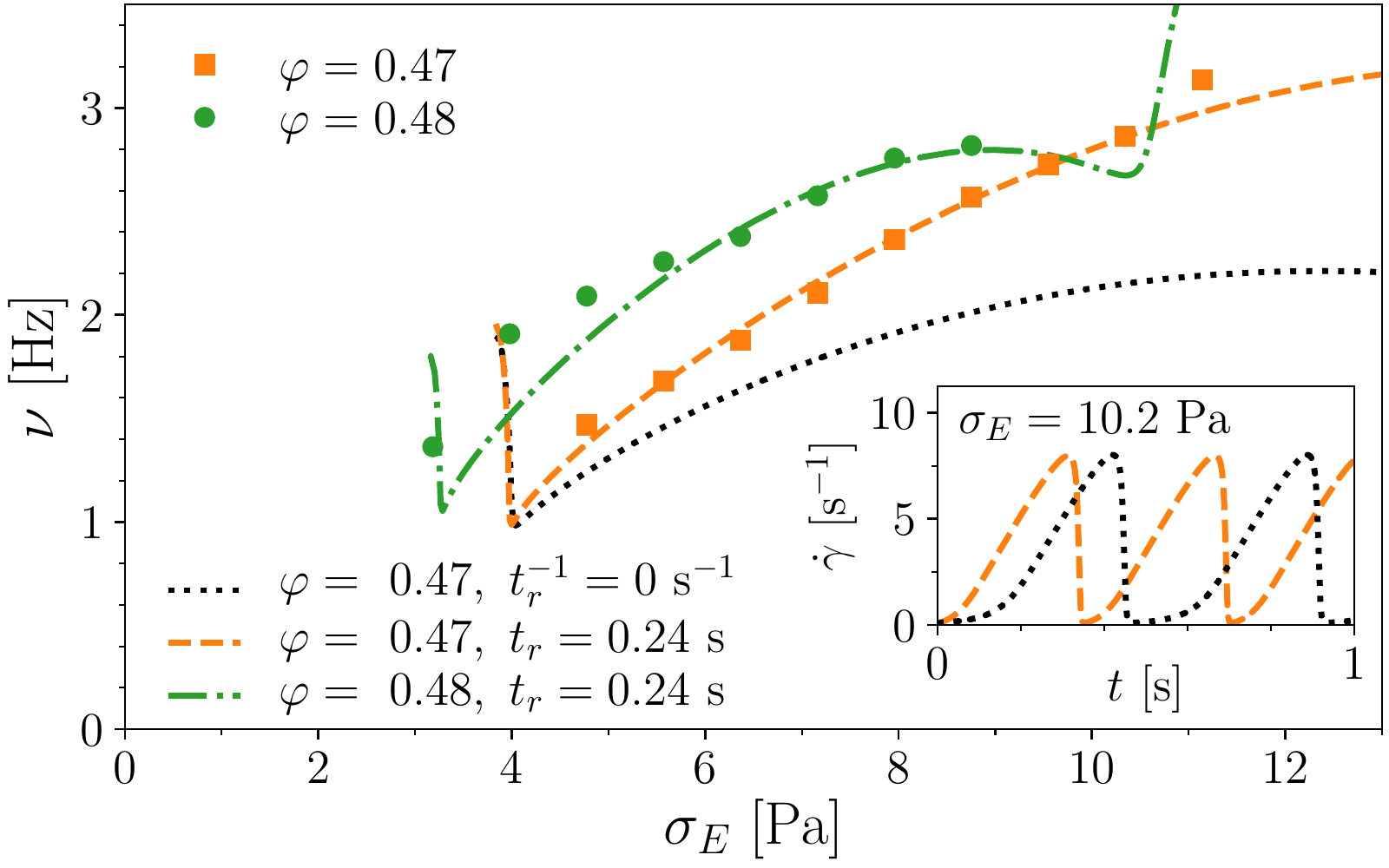}
    \caption{Oscillation frequency, $\nu$, \emph{vs.}~applied stress $\sigma_{E}$. Points: $\nu$ for cornstarch in 50~wt.\% glycerol-water, from Fourier transform of \SI{30}{\second} upwards stress sweep (excluding first \SI{2}{\second}) in steps of \SI{0.796}{\pascal}: $\varphi=0.47~(\begingroup \color{orange} \blacksquare \endgroup),~\varphi=0.48~(\begingroup \color{Green} \bullet \endgroup)$. Lines, model predictions for $\epsilon=2.7\times10^{-6}$ (see legend). Inset: effect of additional time-dependent relaxation on oscillation shape.}
    \label{fig:stress}
\end{figure}

Figure~\ref{fig:trace}(a) pertains to $\sigma_{E}$ at the onset of DST. As $\sigma_{E}$ increases beyond this point, the oscillation frequency, $\nu$, increases~\cite{hermes2016unsteady}, and the agreement between model and experiment worsens, Fig.~\ref{fig:stress}. As the system comes ever closer to jamming at each precipitous drop in $\dot\Gamma$, the strain-dependent \textit{ansatz} for $f$-relaxation, Eq.~\ref{eq:dfdt}, becomes increasingly ineffective. The predicted time taken to traverse $DA$ in the limit cycle, Fig.~\ref{fig:analysis}(d), is lengthened compared to reality (\textit{cf.}~slow onset in Fig.~\ref{fig:stress} inset).

We therefore infer the existence of an additional intrinsic, strain-independent, mechanism for relaxing $f$ towards its steady-state value~\cite{maharjan2017giant} and modify Eq.~\ref{eq:dfdt} to read
\begin{equation}
    \frac { \mathrm{ d } f } { \mathrm { d } t } = - \left(\frac{\dot { \gamma } } { \gamma _ { 0 } } + \frac{1}{t_{r}}\right) \left [ f - \hat { f } ( \eta(f)\dot{\gamma} ) \right ],
    \label{eq:df+time}
\end{equation}
with a new relaxation time $t_{r}$. There are now two contact relaxation mechanisms, dependent on strain ($\propto \dot{\gamma}/\gamma_0$) or time ($\propto 1/t_{r}$). The latter dominates as $\dot\gamma \to 0$, near jamming, so the time taken for $DA$ shortens, decreasing the period of the limit cycle, as observed.

Fitting the $\nu(\sigma_{E})$ data with this new model, Fig.~\ref{fig:stress}~\footnote{The initial frequency spike predicted by the model, associated with small-amplitude limit cycles not seen experimentally~\cite{SM}, is neglected.}, gives $t_{r} \approx$~\SI{0.24 \pm 0.05}{\second}~\cite{SM}. Since $t_{c}/t_{r} = 5 \times 10^{-4} \ll 1$, the strain-dependent mechanism dominates away from jamming~\cite{SM}. Interestingly, $t_{r} \approx$~\SI{0.24}{\second} is comparable to the relaxation time for cornstarch grains pushed into adhesive contact, $\sim$~\SI{0.5}{s}, so that surface chemistry matters~\cite{galvez2017dramatic}.

The mechanism we propose for relaxation oscillations in shear-thickening suspensions, depending on flow-curve shape and geometry inertia, appears generic. It is therefore perhaps a puzzle why such oscillations have not been more widely reported. One reason is the use of high-viscosity solvents, thus giving $\epsilon \gg \epsilon_{c}^{\max}$. More prevalent could be the breakdown of simple shear flow where surface tension no longer confines the particles as sample stress peaks~\cite{brown2012role, guy2015towards} at $C$ in the limit cycle, Fig.~\ref{fig:analysis}(d), causing fracture~\cite{strivens1976shear}. With only two dynamical variables, lacking spatial variation, our model cannot capture such inhomogeneous flow. It nevertheless well captures the development of relaxation oscillations {\it en route} to aperiodic unsteady flows, which {\it are} widely seen~\cite{saint2018uncovering, rathee2017localized, lootens2003giant, pan2015s}.

Our model generalized to Eq.~\ref{eq:df+time} has allowed us to extract an intrinsic contact-relaxation time scale, $t_{r}$, which is difficult to access using other methods such as shear reversal~\cite{lin2015hydrodynamic} or cessation~\cite{maharjan2017giant} due to instrument artifacts. Instead, our method of accessing $t_{r}$ relies on modelling the coupling with one such artifact, viz., geometry inertia. $t_{r}$ becomes important in modelling the flow properties whenever the suspension comes close to jamming and the shear rate drops. With our protocol for extracting this relaxation time, future work should be able to clarify the underlying physical mechanism, which may include particle softness~\cite{ness2016shear}, surface chemistry~\cite{james2018interparticle} or long-range repulsion~\cite{mari2015discontinuous}.

Finally, coupling between fast frictional-contact-network dynamics and a slower `system variable', and hence the resulting types of behavior, should be found in many types of dense suspension flow. Thus, for example, in vorticity banding, particle migration is slow~\cite{chacko2018dynamic}; in micro-channel oscillations, rearrangement due to fluid permeation is slow~\cite{isa2009velocity, kanehl2017self}; in the settling of a ball in a suspension, the ball's inertial dynamics are slow~\cite{von2011nonmonotonic}. Interestingly, relaxation-type oscillations, with periodic bursts of brief near-jamming episodes, have been observed in the pipe flow of polymethylmethacralate particles~\cite{isa2009velocity}, the settling velocity of a ball in cornstarch~\cite{von2011nonmonotonic} and the shear rheology of polystyrene particles~\cite{larsen2014fluctuations} (compare especially data presented in the latter two cases with, {\it e.g.}, our Fig.~\ref{fig:analysis}(c)). It is therefore possible, perhaps likely, that the kind of physics we have modelled may be relevant far beyond the data sets presented here.

The data plotted in this work are available from Edinburgh DataShare~\cite{DataShare}.

\acknowledgments{\textbf{Acknowledgements:} This research was funded by the UK Engineering and Physical Sciences Research Council (EPSRC) [grant numbers EP/J007404/1, EP/N025318/1, EP/L015536/1] and supported in part by the National Science Foundation under Grant No. NSF PHY-1748958 through the KITP program on the Physics of Dense Suspensions. J.A.R. acknowledges funding from the EPSRC Centre for Doctoral Training in Soft Matter and Functional Interfaces (SOFI CDT) and AkzoNobel.}


%

\end{document}